\documentclass[aps,prl,showpacs,showkeys,floatfix,twocolumn,byrevtex,superscriptaddress]{revtex4-1}
\usepackage{amsmath}
\usepackage{amssymb}
\usepackage{amstext}
\usepackage{amsopn}
\usepackage{amsfonts}
\usepackage{amsxtra}
\usepackage[english]{babel}
\usepackage{graphicx}
\usepackage{float}
\usepackage{bm}
\usepackage{multirow}
\usepackage{dcolumn}
\usepackage{color}
\usepackage{hyperref}

\newcommand{\LFCA}{LiFe$_{1-x}$Co$_{x}$As}

\newcommand{\NMRTT}{$1/T_{1}T|_{T = 20\,\text{K}}$}

\begin{document}

\title{\boldmath Spin-fluctuation induced non-Fermi liquid behavior with suppressed superconductivity in LiFe$_{1-x}$Co$_{x}$As \unboldmath}
\author{Y. M. Dai}
\affiliation{Condensed Matter Physics and Materials Science Department, Brookhaven National Laboratory, Upton, New York 11973, USA}
\author{H. Miao}
\author{L. Y. Xing}
\author{X. C. Wang}
\affiliation{Beijing National Laboratory for Condensed Matter Physics, Institute of Physics, Chinese Academy of Sciences, Beijing 100190, China}
\author{P. S. Wang}
\affiliation{Department of Physics, Renmin University of China, Beijing 100872, China}
\author{H. Xiao}
\author{T. Qian}
\affiliation{Beijing National Laboratory for Condensed Matter Physics, Institute of Physics, Chinese Academy of Sciences, Beijing 100190, China}
\author{P. Richard}
\affiliation{Beijing National Laboratory for Condensed Matter Physics, Institute of Physics, Chinese Academy of Sciences, Beijing 100190, China}
\affiliation{Collaborative Innovation Center of Quantum Matter, Beijing, China}
\author{X. G. Qiu}
\affiliation{Beijing National Laboratory for Condensed Matter Physics, Institute of Physics, Chinese Academy of Sciences, Beijing 100190, China}
\author{W. Yu}
\affiliation{Department of Physics, Renmin University of China, Beijing 100872, China}
\author{C. Q. Jin}
\affiliation{Beijing National Laboratory for Condensed Matter Physics, Institute of Physics, Chinese Academy of Sciences, Beijing 100190, China}
\affiliation{Collaborative Innovation Center of Quantum Matter, Beijing, China}
\author{Z. Wang}
\affiliation{Department of Physics, Boston College, Chestnut Hill, Massachusetts 02467, USA}
\author{P. D. Johnson}
\affiliation{Condensed Matter Physics and Materials Science Department, Brookhaven National Laboratory, Upton, New York 11973, USA}
\author{C. C. Homes}
\email[]{homes@bnl.gov}
\affiliation{Condensed Matter Physics and Materials Science Department, Brookhaven National Laboratory, Upton, New York 11973, USA}
\author{H. Ding}
\email[]{dingh@iphy.ac.cn}
\affiliation{Beijing National Laboratory for Condensed Matter Physics, Institute of Physics, Chinese Academy of Sciences, Beijing 100190, China}
\affiliation{Collaborative Innovation Center of Quantum Matter, Beijing, China}

\date{\today}

%
%

\begin{abstract}
A series of LiFe$_{1-x}$Co$_{x}$As compounds with different Co concentrations have been studied by transport, optical spectroscopy, angle-resolved photoemission spectroscopy and nuclear magnetic resonance. We observed a Fermi liquid to non-Fermi liquid to Fermi liquid (FL-NFL-FL) crossover alongside a monotonic suppression of the superconductivity with increasing Co content. In parallel to the FL-NFL-FL crossover, we found that both the low-energy spin fluctuations and Fermi surface nesting are enhanced and then diminished, strongly suggesting that the NFL behavior in LiFe$_{1-x}$Co$_{x}$As is induced by low-energy spin fluctuations which are very likely tuned by Fermi surface nesting. Our study reveals a unique phase diagram of LiFe$_{1-x}$Co$_{x}$As where the region of NFL is moved to the boundary of the superconducting phase, implying that they are probably governed by different mechanisms.
\end{abstract}


\pacs{74.70.Xa,74.25.Dw,71.10.Hf}
\keywords{Condensed Matter Physics, Superconductivity}

\maketitle



The normal state of high-temperature (high-$T_c$) superconductors is very unusual,
with the electrical resistivity (or quasiparticle scattering rate) varying with temperature in a peculiar way
that deviates significantly from the quadratic $T$ dependence expected from Landau's Fermi-liquid (FL) theory of
metals~\cite{Boebinger2009,Cooper2009,Jin2011,Lohneysen2007}. Because this anomalous non-Fermi-liquid
(NFL) behavior is often revealed experimentally above a superconducting dome, there is a broad consensus
that the origin of the NFL behavior may hold the key to understanding the pairing mechanism of high-$T_c$
superconductivity.

Studies on high-$T_{c}$ cuprate superconductors~\cite{Boebinger2009,Cooper2009,Jin2011}, heavy-fermion
metals~\cite{Lohneysen2007,Gegenwart2008}, organic Bechgaard salts~\cite{Doiron-Leyraud2009} as well as
the newly-discovered iron-based superconductors (IBSCs)~\cite{Hashimoto2012,Zhou2013,Analytis2014,Shibauchi2014}
have shown that the NFL behavior and high-$T_c$ superconducting dome favor proximity to magnetic order.
This fact has led to proposals ascribing both the NFL behavior and high-$T_c$ superconductivity to
spin fluctuations close to a magnetic quantum critical point (QCP)~\cite{Moriya2000,Taillefer2010,Sachdev2011}.
However, a growing number of experiments do not agree with these scenarios. For example, a recent
magnetotransport study has shown that by doping CeCoIn$_{5}$ with Yb, the field-induced QCP is fully
suppressed while both the NFL behavior and superconductivity are barely affected~\cite{Hu2013}.
At the current time, the microscopic mechanism of the NFL behavior and its relationship to high-$T_c$
superconductivity are still a matter of considerable debate.

IBSCs feature an intricate phase diagram with NFL behavior, superconducting phase, magnetic order, structural transition, possible QCP(s) and nested Fermi surfaces interacting with each other. This complexity makes it quite challenging to distinguish the roles played by different orders or interactions. The \LFCA\ system presents a simple phase diagram: LiFeAs exhibits superconductivity with a maximum transition temperature $T_{c} \approx 18$~K in its stoichiometric form~\cite{Wang2008}. The substitution of Fe by Co results in a monotonic lowering of $T_{c}$; neither magnetic nor structural transitions have been detected in the temperature--doping ($T$--$x$) phase diagram of \LFCA~\cite{Pitcher2010}. The normal state of LiFeAs is a FL, as evidenced by the quadratic $T$
dependence of the low-temperature resistivity~\cite{Heyer2011,Albenque2012}. Such a simple phase diagram makes \LFCA\ an excellent system to elucidate the origin of the NFL behavior and its relationship to superconductivity.

In this article, through a combined study of transport, optical spectroscopy, angle-resolved photoemission spectroscopy (ARPES) and nuclear magnetic resonance (NMR) on \LFCA, we found that while superconductivity is monotonically suppressed with increasing Co concentration, the transport and optical properties reveal a prominent FL-NFL-FL crossover which closely follows the doping evolution of low-energy spin fluctuations (LESFs) and Fermi surface nesting. Our observations provide clear evidence that LESFs, which are likely tuned by FS nesting, dominate the normal-state scattering, and are thus responsible for the FL-NFL-FL crossover in \LFCA. A unique phase diagram of \LFCA\ derived from our studies shows that the NFL behavior is decoupled from superconductivity, suggesting that they do not share the same origin.

High quality single crystals of \LFCA\ with different Co concentrations were grown by a self-flux method~\cite{Wang2008}. Details of the sample synthesis and experimental methods for all the techniques we used in this work are included in Appendixes A to E.

%
\begin{figure}[tb]
\includegraphics[width=\columnwidth]{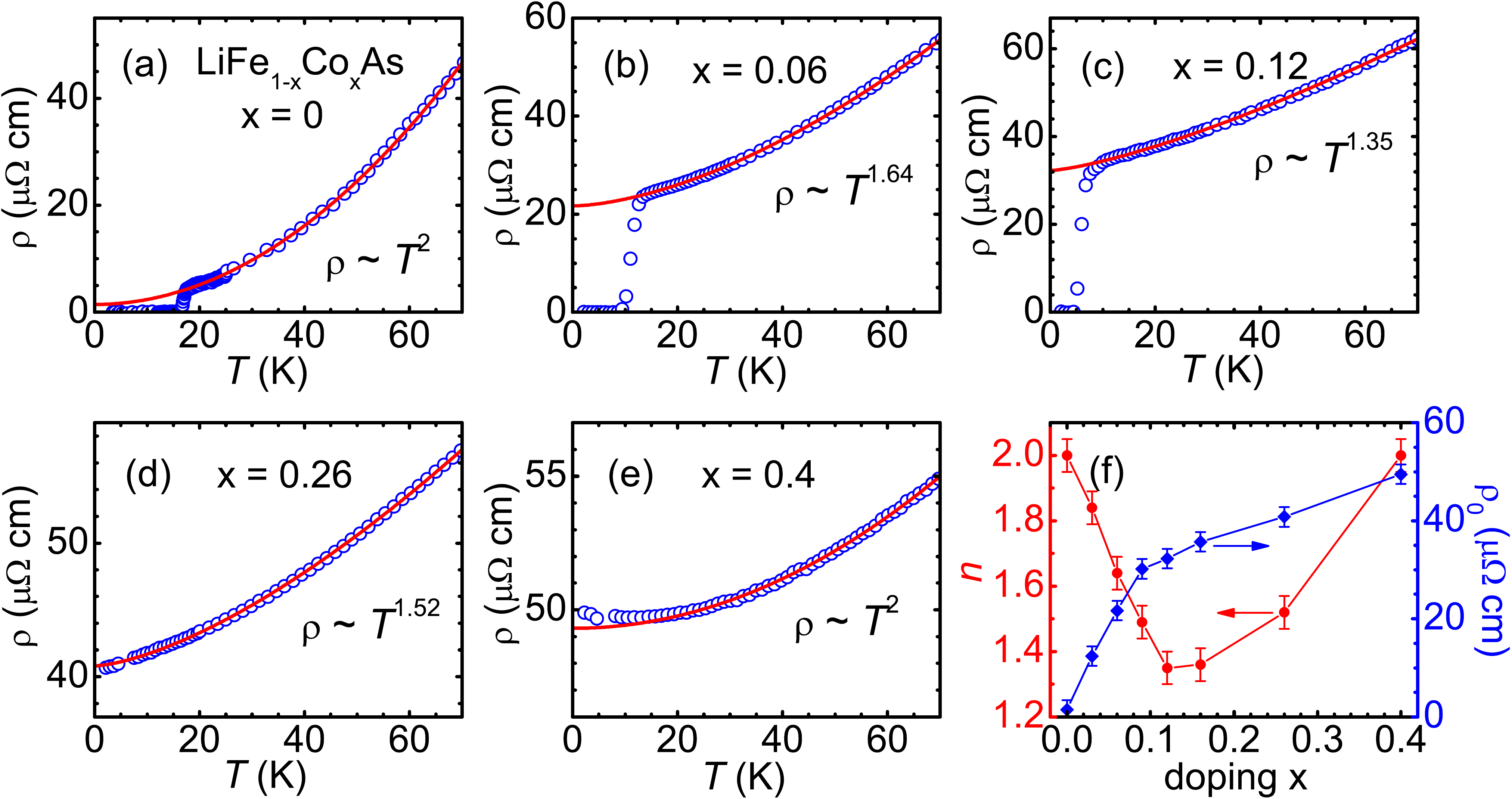}
\caption{ (color online) (a)--(e) Resistivity as a function of temperature $\rho(T)$ (open circles) for LiFe$_{1-x}$Co$_{x}$As at five selected Co concentrations. For each material, $\rho(T)$ is fit to a single power law $\rho(T) = \rho_{0} + A T^{n}$ (solid lines). (f) Evolution of the exponent $n$ (solid circles) and residual resistivity $\rho_{0}$ (solid diamonds) with Co substitution $x$.}
\label{Fig1}
\end{figure}
Figures~\ref{Fig1}(a)--\ref{Fig1}(e) show the $T$-dependent resistivity $\rho(T)$ for five representative dopings. For each doping, $\rho(T)$ is fit to a single power law $\rho(T) = \rho_{0} + A T^{n}$, returning the exponent $n$ and the residual resistivity $\rho_{0}$. The evolution of $n$ and $\rho_{0}$ with doping are summarized in Fig.~\ref{Fig1}(f) as solid circles and solid diamonds, respectively. In LiFeAs ($x = 0$), $n = 2$, \emph{i.e.} the resistivity varies quadratically with temperature, indicating a FL normal state, in agreement with previous transport studies on LiFeAs~\cite{Heyer2011,Albenque2012}. With increasing Co doping, $n$ decreases, reaching a minimum of 1.35 at $x \approx 0.12$. With further doping ($x > 0.12$), $n$ begins to increase and recovers to a value of 2 again at about $x = 0.4$. $\rho(T)$ and the single power-law fitting results for more dopings are displayed in Appendix B (Fig.~\ref{FigSRT}). To present the crossover behavior more clearly, $\rho(T)$ is plotted as a function of $T^{n}$, as shown in Appendix B (Fig.~\ref{FigSRTn}), where a linear behavior can be seen for all the dopings. This doping dependence of $n$ is an explicit indication of a doping-induced FL-NFL-FL crossover in \LFCA. In addition, we note that, as shown in Fig.~\ref{Fig1}(f), $\rho_{0}$ increases with doping all the way to $x = 0.4$, while $n$ exhibits a FL-NFL-FL crossover in the same doping range. This indicates that the FL-NFL-FL crossover in \LFCA\ is not tied to the impurity level.

%
\begin{figure}[tb]
\includegraphics[width=\columnwidth]{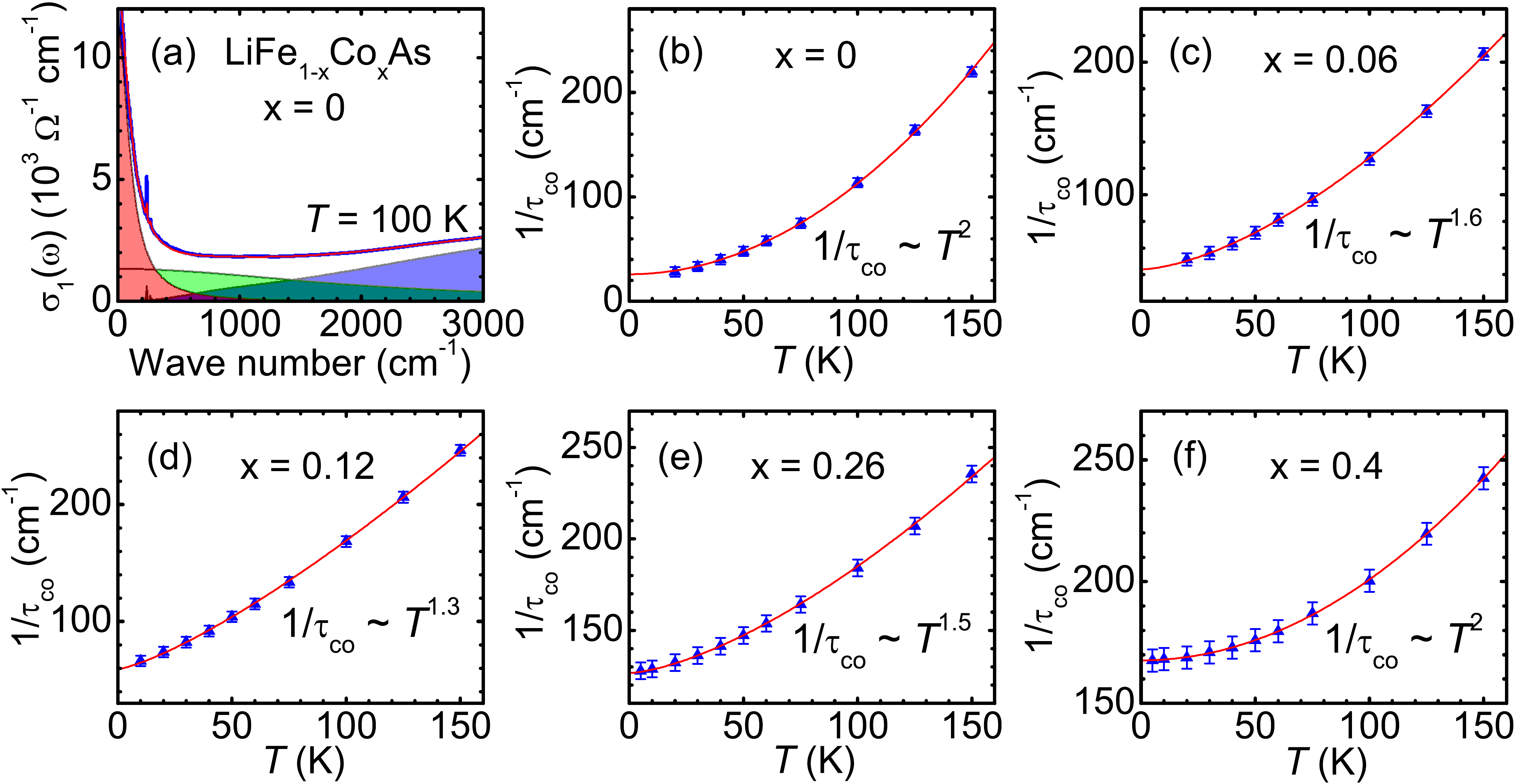}
\caption{ (color online) (a) The thick blue curve is the real part of the optical conductivity
$\sigma_{1}(\omega)$ of LiFeAs ($x=0$) measured at 100~K. The thin red curve through the data is the
Drude-Lorentz fit which consists of the contributions from a coherent narrow Drude (red shaded region), a
nearly incoherent broad Drude (green shaded region) and series of Lorentz components (blue shaded region).
(b)--(f) $T$ dependence of the quasiparticle scattering rate $1/\tau_{\rm co}$
derived from the coherent narrow Drude component for five Co concentrations.}
\label{Fig2}
\end{figure}
Further evidence for the FL-NFL-FL crossover can be revealed by the $T$ dependence of the quasiparticle
scattering rate obtained via optical spectroscopy~\cite{Wu2010a,Dai2013}.
Figure~\ref{Fig2}(a) displays the real part of the optical conductivity $\sigma_{1}(\omega)$ for LiFeAs ($x=0$) at 100~K. The low-frequency
$\sigma_{1}(\omega)$ is dominated by the well-known Drude-like metallic response, where the width of
the Drude peak at half maximum gives the value of the quasiparticle scattering rate. In order to
accurately extract the quasiparticle scattering rate, we fit the measured $\sigma_{1}(\omega)$ to the
Drude-Lorentz model:
%
%
\begin{equation}
\sigma_{1}(\omega) = \frac{2\pi}{Z_{0}} \left[
   \sum_{k} \frac{\omega^{2}_{p,k}}{\tau_{k}(\omega^{2}+\tau_{k}^{-2})} +
   \sum_{j} \frac{\gamma_{j} \omega^{2} \Omega_{j}^{2}}{(\omega_{j}^{2} - \omega^{2})^{2} + \gamma_{j}^{2} \omega^{2}}
 \right],
\label{DLModel}
\end{equation}
where $Z_{0} \simeq 377\,\Omega$ is the impedance of free space. The first term describes a sum of
delocalized (Drude) carrier responses with $\omega_{p,k}$ and $1/\tau_{k}$ being the plasma frequency
and scattering rate in the $k$th Drude band, respectively. In the second term, $\omega_{j}$, $\gamma_{j}$
and $\Omega_{j}$ are the resonance frequency, width and strength of the $j$th vibration or bound
excitation. As shown in Fig.~\ref{Fig2}(a), the thin red line through the data represents the fitting result for LiFeAs at 100~K which is decomposed into a coherent narrow Drude, a nearly incoherent broad Drude, and series of Lorentz components, consistent with previous optical studies on IBSCs~\cite{Wu2010a,Dai2013,Tu2010,Nakajima2014}. Fitting results for other dopings at several representative temperatures can be found in Appendix C [Fig.~\ref{FigSOpt}(f)--\ref{FigSOpt}(j)].
%
%
Tu \emph{et al.} suggest that it is more appropriate to describe the broad Drude component as
bound excitations~\cite{Tu2010} because the mean free path $l = v_{F} \tau$ ($v_{F}$ is the Fermi velocity)
associated with the broad Drude component is close to the Mott-Ioffe-Regel limit. In any event, since the broad
Drude component only gives rise to a $T$-independent background contribution to the total $\sigma_{1}(\omega)$,
the $T$ dependence of the optical response is governed by the coherent narrow Drude component. As a result, the nature
of the broad Drude term does not affect our analysis of the coherent narrow Drude component and the $T$ dependence of
$\sigma_{1}(\omega)$.
The application of the Drude-Lorentz analysis at all the measured temperatures for five representative
dopings yields the $T$ dependence of the scattering rate of the coherent narrow Drude component $1/\tau_{\rm co}$,
shown in Fig.~\ref{Fig2}(b)--\ref{Fig2}(f). For each doping, $1/\tau_{\rm co}$ follows the expression $1/\tau_{\rm co} =
1/\tau_{0} + B T^{\alpha}$ with the exponent $\alpha \approx n$, where $n$ is the exponent determined
from the fit to $\rho(T)$ for the corresponding dopings. Again, we plot $1/\tau_{\rm co}$ as a function of
$T^{\alpha}$ in Appendix C [Fig.~\ref{FigSOpt}(k)--\ref{FigSOpt}(o)], which reveals distinct linear behavior for all the dopings. Such
crossover behavior of $\alpha$ provides further evidence for the doping-induced FL-NFL-FL changes
in \LFCA.


In order to gain insight into the origin of the anomalous FL-NFL-FL crossover in \LFCA, we examine the evolution of LESFs with Co concentration by looking into the nuclear spin-lattice relaxation rate $1/T_{1}T$, reflecting the summation of all different \textbf{q} modes of sub-meV LESFs weighted by a nearly uniform form factor, which can be determined from $^{75}$As nuclear magnetic resonance (NMR) measurements~\cite{Nakai2010,Zhou2013}.
%
\begin{figure}[tb]
\includegraphics[width=\columnwidth]{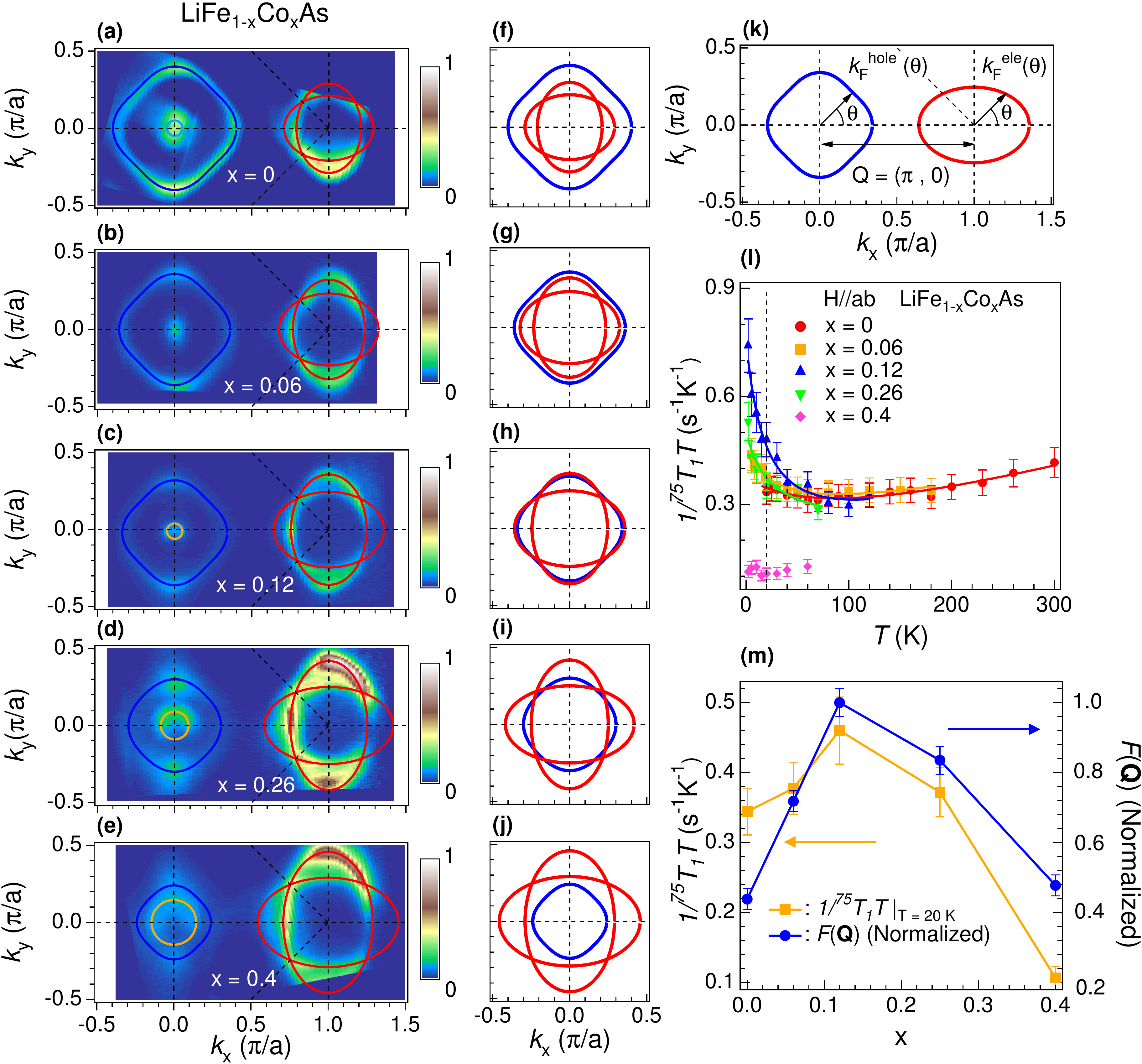}
\caption{ (color online) (a)--(e) FS contour of \LFCA\ for five representative Co concentrations,
determined by integrating the ARPES spectral intensity within $\pm$10 meV with respect to $E_F$.
(f)--(j) Extracted FSs from corresponding ARPES measurements for each doping.
(k) Definition of $\mathbf{k}_\mathbf{F}^{hole}(\theta)$, $\mathbf{k}_\mathbf{F}^{ele}(\theta)$
and $\mathbf{Q}$ in the $k$-space.
(l) Spin-lattice relaxation rate $^{75}{\rm As}$ $1/T_{1}T$ as a function of temperature
for five Co concentrations measured by NMR.
(m) Evolution of $^{75}{\rm As}$ $1/T_{1}T$ at 20~K (solid squares)
and FS nesting factor (solid circles) with increasing Co concentration.}
\label{Fig3}
\end{figure}
Figure~\ref{Fig3}(l) displays the $T$ dependence of $1/T_{1}T$ for five different dopings.  In LiFeAs ($x = 0$), $1/T_{1}T$ is almost $T$ independent, and a negligible upturn at low temperature indicates weak LESFs.  However, in sharp contrast to LiFeAs, $1/T_{1}T$ for $x = 0.12$ increases rapidly upon cooling and a prominent upturn develops at low temperature, implying that LESFs are significantly enhanced. For $x = 0.26$, the upturn in $1/T_{1}T$ at low temperature becomes less prominent, suggesting that the LESFs diminish again. The low-temperature upturn in $1/T_{1}T$ can be empirically described by the Curie-Weiss expression $1/T_{1}T = A + BT + C/(T + \theta)$ [solid lines in Fig.~\ref{Fig3}(l)], in good agreement with previous NMR studies~\cite{Nakai2010,Zhou2013}. To quantify the doping dependence of the LESFs, we take the value of $1/T_{1}T$ at 20~K (just above $T_c$ of LiFeAs), \NMRTT, for each doping and plot them as a function of $x$ [solid squares in Fig.~\ref{Fig3}(m)]. Upon doping, \NMRTT\ first grows but then drops, resulting in a peak at $x = 0.12$. The link between the LESFs and the NFL behavior can be revealed by comparing the doping dependence of $n$ [solid circles in Fig.~\ref{Fig1}(f)] and \NMRTT\ [solid squares in Fig.~\ref{Fig3}(m)].  Below $x = 0.12$, the enhancement of LESFs (increase in \NMRTT ) leads to a more conspicuous deviation from a FL (decrease in $n$) while above $x = 0.12$, the reduction of LESFs (decrease in \NMRTT) results in a gradual recovery of the FL behavior (increase in $n$). The most robust NFL behavior ($n \approx 1.35$) occurs at $x = 0.12$ where the LESFs are optimized (\NMRTT\ peaks). These observations strongly suggest that the FL-NFL-FL crossover in \LFCA\ is governed by LESFs.

Since an investigation into the FS may provide information on the nature of the LESFs, we then studied the evolution of the FSs in \LFCA. The FS contour of \LFCA\ is traced out from the ARPES intensity plot near the Fermi energy ($E_{F}$) for five representative dopings [Fig.~\ref{Fig3}(a)--\ref{Fig3}(e)].  The extracted FSs for each doping are shown in Fig.~\ref{Fig3}(f)--\ref{Fig3}(j).  For LiFeAs ($x = 0$), two hole and two electron FS pockets are observed at the $\Gamma$ and $M$ points, respectively [Fig.~\ref{Fig3}(a)], in accord with previous ARPES studies~\cite{Borisenko2010,Umezawa2012}. The inner hole pocket is quite small, while the outer hole pocket is much larger than the electron pockets, resulting in a poor nesting condition in LiFeAs [Fig.~\ref{Fig3}(f)]. With Co doping, the electron pockets expand while the hole pockets shrink.  Consequently, the FS nesting is improved. As shown in Fig.~\ref{Fig3}(h), the shape of the hole FS matches the outer contour of the two electron FSs at $x \approx 0.12$. Further Co doping ($x > 0.12$) makes the electron and hole pockets mismatched again, \emph{e.g.} $x = 0.4$ [Fig.~\ref{Fig3}(e) and \ref{Fig3}(j)], leading to a degradation of the FS nesting. In order to quantitatively analyze the FS nesting, we define a nesting factor at vector $\mathbf{Q}$
%
%
\begin{equation}
F(\mathbf{Q}) = \sum_{i,j} \int_{0}^{2\pi} \frac{1}{\| \mathbf{k}_\mathbf{F}^{ele_{i}}(\theta) - \mathbf{k}_\mathbf{F}^{hole_{j}}(\theta) -
    \mathbf{Q} \|+\delta} d\theta
\label{FSNesting}
\end{equation}
where the definitions of $\mathbf{k_{F}}^{ele}(\theta)$, $\mathbf{k_{F}}^{hole}(\theta)$ and $\mathbf{Q}$ are illustrated in Fig.~\ref{Fig3}(k); $\delta$ is a small positive number to avoid singular behavior; for an $n$-dimensional vector $\mathbf{x}$, $||\mathbf{x}|| = \sqrt{\mathbf{x} \cdot \mathbf{x}} = \sqrt{\sum_{i=1}^{n}(x_{i})^{2}}$. $F(\mathbf{Q})$ increases as the FS nesting is improved, and is maximized when the hole FS matches the outer contour of the two electron FSs. Assuming $v_{F}$ is uniform on all FSs, the nesting factor $F(\mathbf{Q})$ is proportional to the non-interacting single-orbital magnetic susceptibility at vector $\mathbf{Q}$:
%
%
\begin{equation}
  \chi^{(0)}(\mathbf{Q}) = \sum_{\mathbf{p}} \frac{f_{\mathbf{p}} - f_{\mathbf{p} + \mathbf{Q}}}{\varepsilon_{\mathbf{p} + \mathbf{Q}} - \varepsilon_{\mathbf{p}}}
\label{SOMSus}
\end{equation}
where $f_{\mathbf{p}}$ is the Fermi distribution and $\varepsilon_{\mathbf{p}}$ is the quasiparticle kinetic energy. $F(\mathbf{Q})$ is calculated by Eq.~(\ref{FSNesting}) for each doping and normalized by its value at $x = 0.12$. All FSs have been considered in the calculation. The solid circles in Fig.~\ref{Fig3}(m) portray the doping dependence of $F(\mathbf{Q})$. Remarkably, $F(\mathbf{Q})$ follows exactly the same doping dependence as \NMRTT, indicating that the LESFs probed by NMR are closely related to the FS nesting. The FS structure naturally suggests that the spin fluctuations are of the antiferromagnetic type with large wave vectors close to the nesting vectors $(\pm{\pi},0)$ and $(0,\pm{\pi})$. This is indeed consistent with our NMR data.  The Knight shift that measures the uniform susceptibility becomes $T$ independent below 30~K (Appendix E, Fig.~\ref{FigSNMR}), indicating that the low-temperature upturn in $1/T_1T$ comes from large-momentum spin fluctuations. Note that since the NMR form factor for As is known to be broadly distributed in momentum space in IBSCs~\cite{Kitagawa2010}, both commensurate (close to FS nesting) and incommensurate (away from nesting) LESFs are captured by the spin-lattice relaxation rate $1/T_{1}T$.

%
\begin{figure}[tb]
\includegraphics[width=\columnwidth]{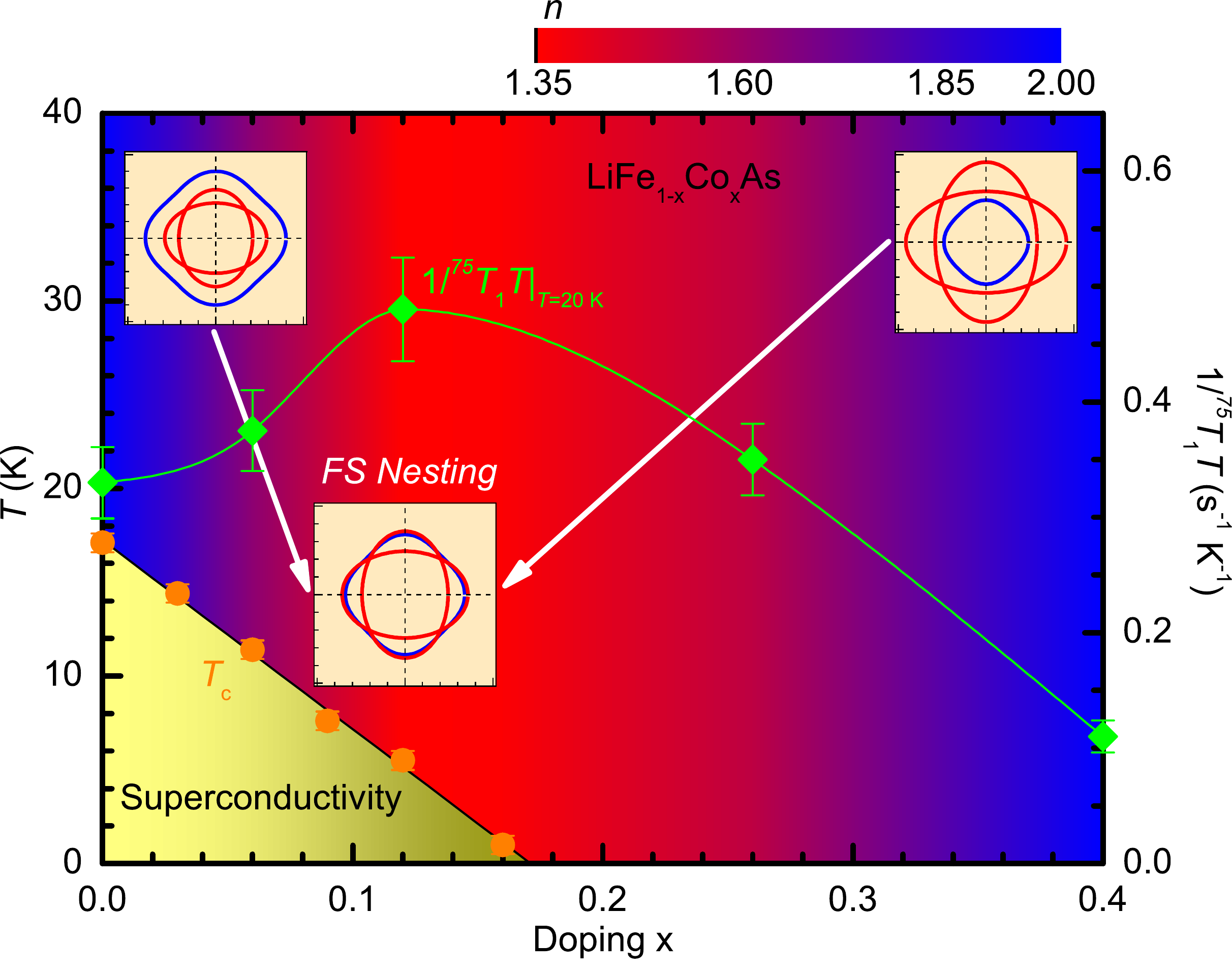}
\caption{ (color online) Temperature--doping ($T$--$x$) phase diagram of \LFCA. Superconductivity (yellow regime)
is monotonically suppressed with increasing Co concentration $x$ and terminates at a critical value
$x \approx 0.17$.  The normal state of LiFeAs is a Fermi liquid (blue regime at $x = 0$), where the
$T$-dependent resistivity follows $\rho(T) = \rho_{0} + A T^{n}$ with $n = 2$.  A crossover from Fermi liquid
to non-Fermi liquid is induced by Co doping.  At $x = 0.12$, $n$ reaches its minimum value of $1.35$, indicating
the most robust non-Fermi liquid behavior (red regime).  Further doping results in a reversal of this trend
until by $x = 0.4$, the Fermi liquid behavior is fully recovered (blue regime at $x = 0.4$). The green diamonds
denote the spin-lattice relaxation rate at 20~K (\NMRTT) measured by NMR for several representative dopings.
\NMRTT\ reaches the maximum at $x \approx 0.12$, signifying that low-energy spin fluctuations are optimized
at this doping.  The three inset panels depict the extracted Fermi surfaces for three representative
Co concentrations: $x = 0$ (left), $x = 0.12$ (middle) and $x = 0.4$ (right).  While the Fermi surface
nesting is poor for $x=0$ and $0.4$, the nesting condition is significantly improved at $x = 0.12$.}
\label{Fig4}
\end{figure}
In Fig.~\ref{Fig4} we summarize our experimental results in the $T$--$x$ phase diagram of \LFCA.  With increasing
Co concentration $x$ that monotonically suppresses the superconducting transition temperature $T_c$
by electron-doping:
(i) The $T$ dependence of the resistivity $\rho(T) \propto A T^n$ and the optical scattering rate
$1/\tau(T) \propto B T^\alpha$ deviates from a FL ($n,\alpha=2$) observed near $x=0$, reaching the
most pronounced NFL power-law behavior ($n \simeq \alpha \simeq 1.35$) at $x \simeq 0.12$ and
then gradually returns for $x > 0.12$ to the FL values at $x=0.4$;
(ii) The NMR spin-lattice relaxation rate $1/T_1T$ shows that LESFs, small at $x=0$, gradually enhance
and become strongest at $x = 0.12$, but diminish for $x > 0.12$;
(iii) ARPES measurements reveal that while the electron and hole FS pockets are far from being nested
at $x = 0$, the nesting improves with doping, is optimized near $x = 0.12$, and then degrades with
further electron doping for $x > 0.12$;
(iv) No long range magnetic order is observed in the $T$--$x$ phase diagram of \LFCA\ up to $x = 0.4$, which is consistent with previous studies~\cite{Wang2008,Pitcher2010,Chu2009b,Qureshi2012}.

A comparison between the above observations (i) and (ii) strongly suggests that the FL-NFL-FL crossover in \LFCA\ is induced by LESFs. Point (iii) in combination with (i) and (ii) implies that the integrated LESFs probed by NMR are dominated by, or at least scale with those near $\mathbf{q} \sim \mathbf{Q}$ which are most likely tuned by FS nesting. The fact (iv) does not directly support a magnetic QCP in the $T$--$x$ phase diagram of \LFCA. However, the pronounced NFL behavior observed near $x = 0.12$, which is usually considered as a signature of quantum criticality~\cite{Cooper2009,Gegenwart2008,Sachdev2011}, in conjunction with the strong tendency to diverge in $1/T_{1}T$ upon cooling at the same doping, points to an incipient QCP near $x = 0.12$. A magnetic order may emerge under pressure or magnetic field in the material with $x = 0.12$, resulting in an actual magnetic QCP associated with other tuning parameters. Finally, the NFL behavior is observed at the boundary of the superconducting phase, implying that they are likely to be governed by different mechanisms.

%
%

Y. M. Dai, H. Miao and L. Y. Xing contributed equally to this work. We thank A. Akrap, S. Biermann, P. C. Dai, J. P. Hu, W. Ku, R. P. S. M. Lobo, A. J. Millis, A. van Roekeghem, W. G. Yin, I. Zaliznyak and G. Q. Zheng for valuable discussion.  Work at BNL was supported by the U.S. Department of Energy, Office of Basic Energy Sciences, Division of Materials Sciences and Engineering under Contract No. DE-SC0012704.  Work at IOP was supported by grants from CAS (XDB07000000), MOST (2010CB923000, 2011CBA001000 and 2013CB921700), NSFC (11234014, 11274362, 11220101003 and 11474344). Work at RUC was supported by the National Basic Research Program of China under Grants No. 2010CB923004 and 2011CBA00112 and by the NSF of China under Grants No. 11222433 and 11374364. Work at BC was supported by U.S. Department of Energy, Office of Science, Basic Energy Sciences, under Award DE-FG02-99ER45747.

%
%

\appendix

\section{APPENDIX A: SAMPLE SYNTHESIS AND CHARACTERIZATION}

\noindent\textbf{Sample synthesis}\hspace{2mm}
High-quality single crystals of \LFCA\ were grown with the self-flux method. The precursor of
Li$_{3}$As was prepared by sintering Li foil and an As lump at about 700~$^{\circ}$C for
10~h in a Ti tube filled with argon (Ar) atmosphere. Fe$_{1-x}$Co$_{x}$As was prepared by mixing
the Fe, Co and As powders thoroughly, and then sealed in an evacuated quartz tube, and
sintered at 700~$^{\circ}$C for 30~h. To ensure the homogeneity of the product, these
pellets were reground and heated for a second time.  The Li$_{3}$As, Fe$_{1-x}$Co$_{x}$As,
and As powders were mixed according to the elemental ratio Li(Fe$_{1-x}$Co$_{x}$)$_{0.3}$As.
The mixture was put into an alumina oxide tube and subsequently sealed in a Nb tube and
placed in a quartz tube under vacuum. The sample was heated at 650~$^{\circ}$C for 10~h and
then heated up to 1000~$^{\circ}$C for another 10~h. Finally, it was cooled down to 750~$^{\circ}$C
at a rate of 2~$^{\circ}$C per hour.  Crystals with a size up to 5~mm were obtained. The entire
process of preparing the starting materials and the evaluation of the final products were carried
out in a glove box purged with high-purity Ar gas. \\

\noindent\textbf{Determination of the doping level}\hspace{2mm}
The molar ratio of Co and Fe of the \LFCA\ single crystals was checked by energy dispersive x-ray
spectroscopy (EDS) at several points on one or two selected samples for each Co concentration.
For each doping, the Co concentration measured by EDS is consistent with the nominal value.

%
\begin{figure*}[tb]
\includegraphics[width=0.95\textwidth]{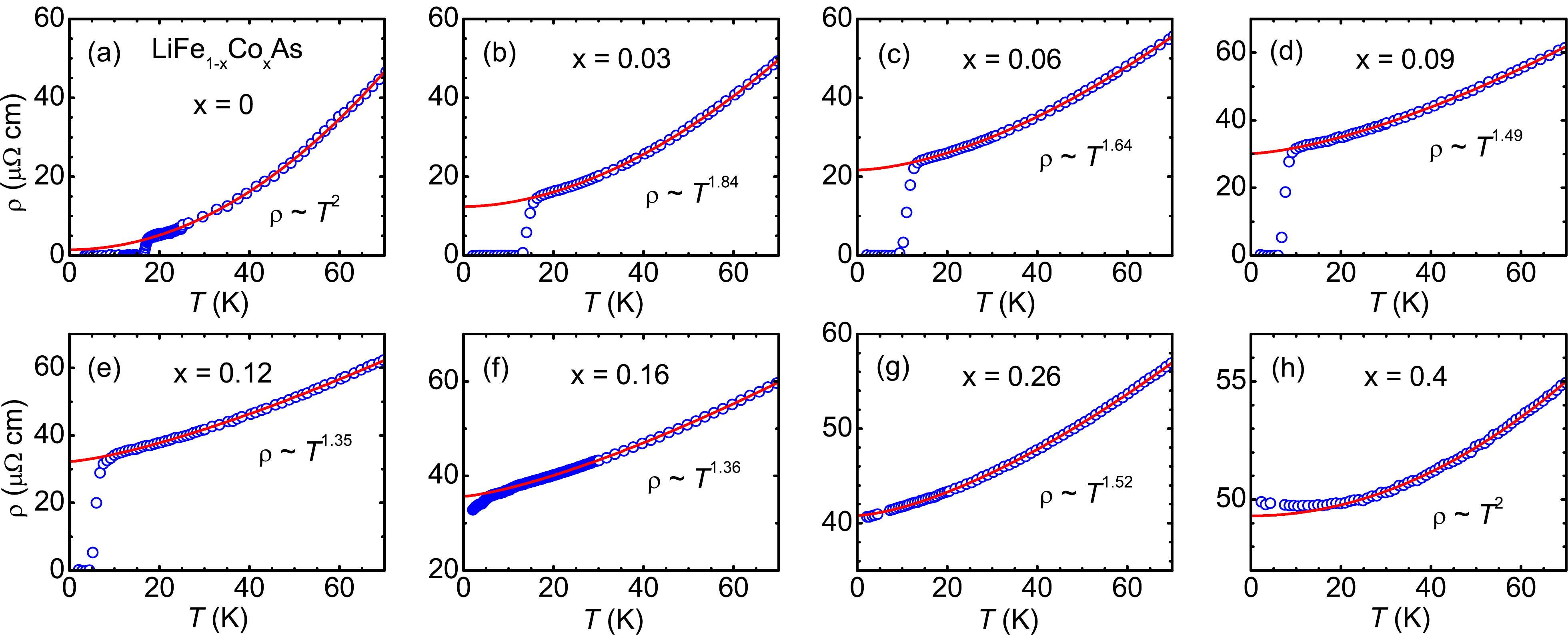}
\caption{ (color online) Resistivity of LiFe$_{1-x}$Co$_x$As as a function of temperature for all the Co values.
For each sample, the resistivity curve is fit to the
single power law expression $\rho(T)=\rho_0+AT^n$.  The open circles denote the
measured resistivity and the solid lines in each panel are the fitting results. The power
$n$ derived from the fitting is shown in each panel. }
\label{FigSRT}
\end{figure*}
%
\begin{figure*}[tb]
\includegraphics[width=0.95\textwidth]{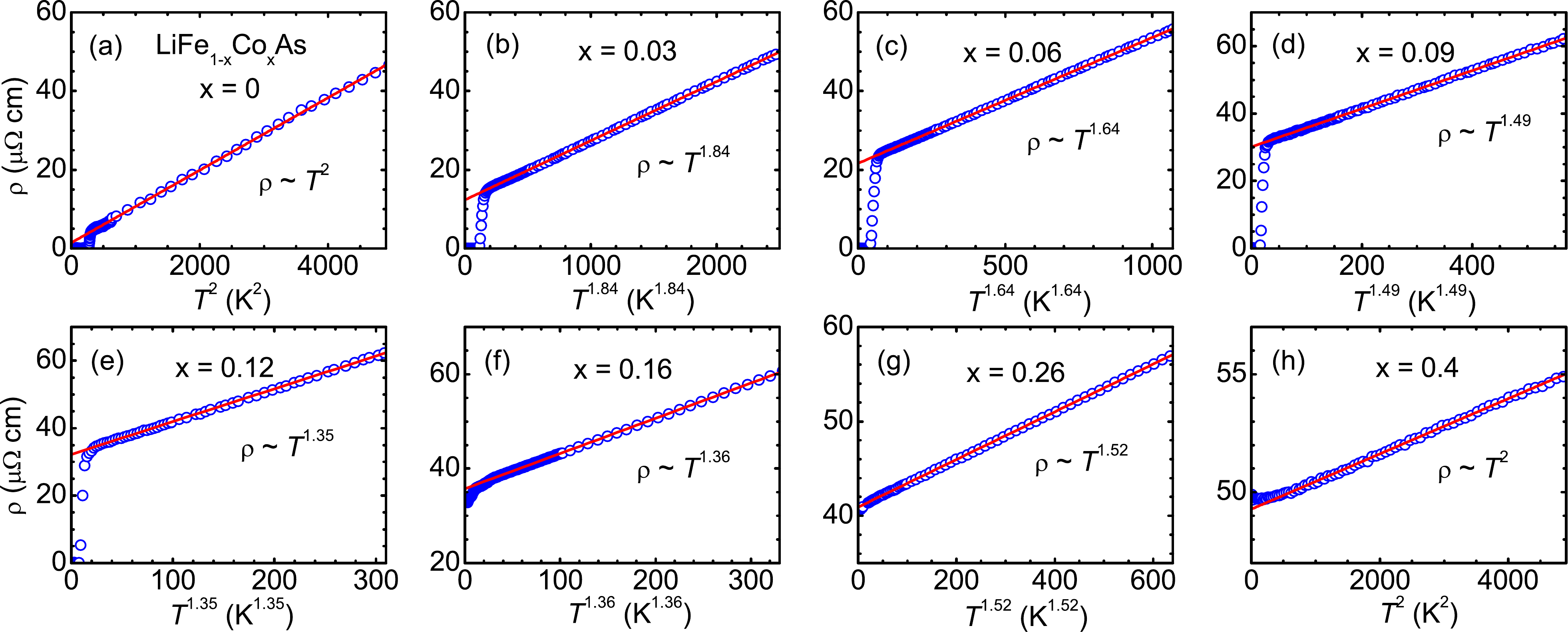}
\caption{ (color online) Resistivity of LiFe$_{1-x}$Co$_x$As
  as a function of $T^n$ for all the Co values.
$n$ is the power determined from the single power law fit to the resistivity as a
function of $T$. The straight solid line in each panel is a guide to the eye.}
\label{FigSRTn}
\end{figure*}
%
\begin{figure*}[tb]
\includegraphics[width=0.95\textwidth]{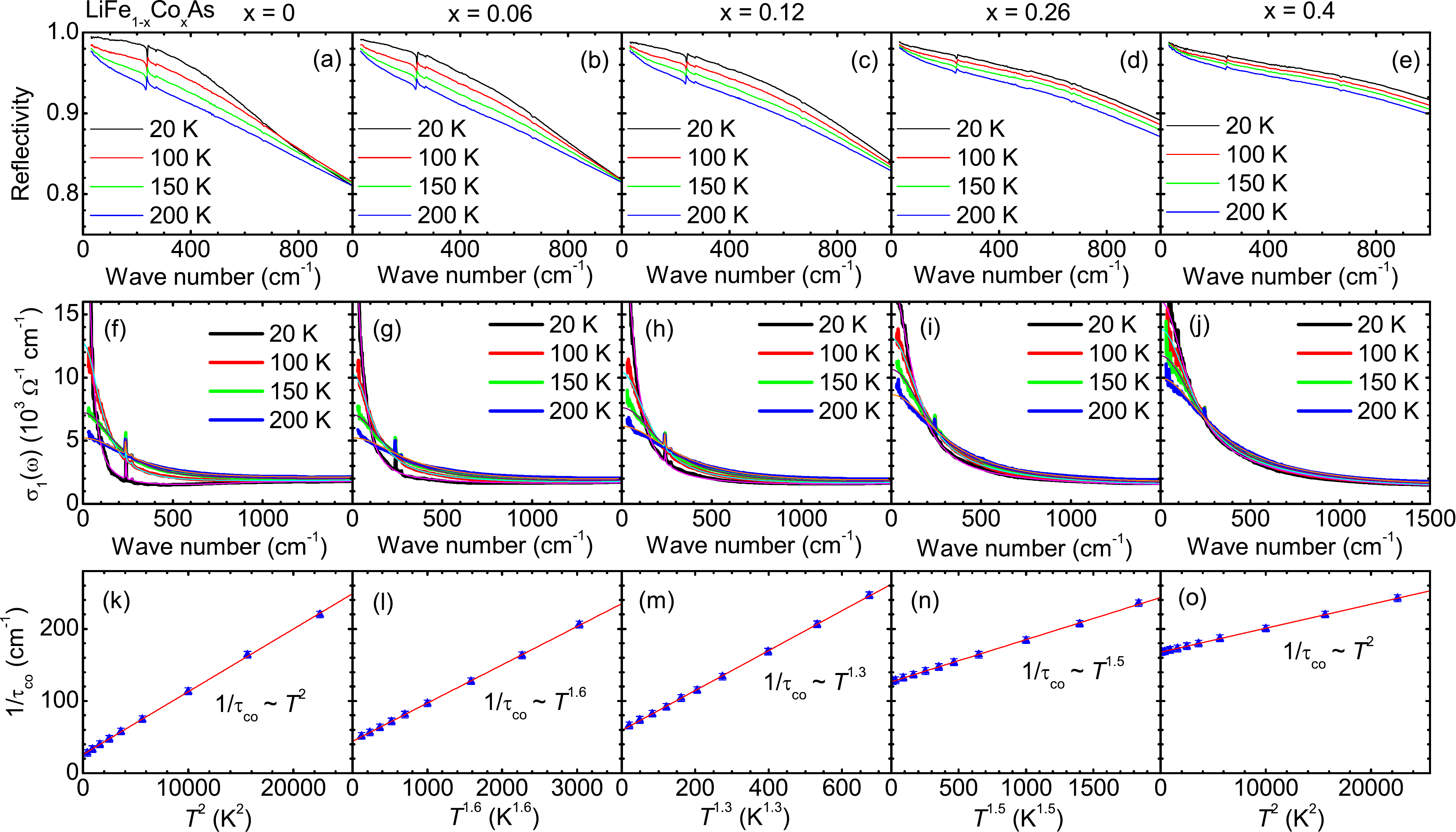}
\caption{ (color online) Reflectivity, optical conductivity and
 quasiparticle scattering rate of \LFCA.
(a)--(e) Temperature dependence of the reflectivity of \LFCA\ in the
far-infrared region at several temperatures for the stoichiometric material and 4 representative
Co concentrations. (f)--(j) Real part of the optical conductivity
derived from the reflectivity. The thick solid curves are the experimental data and the
thin solid curves through the data denote the Drude-Lorentz fitting results.
(k)--(o) Scattering rate of the coherent narrow Drude component ($1/\tau_{\rm co}$)
as a function of $T^{\alpha}$. The straight solid line in each panel is a guide to the eye.}
\label{FigSOpt}
\end{figure*}
%
\begin{figure}[tb]
\includegraphics[width=0.95\columnwidth]{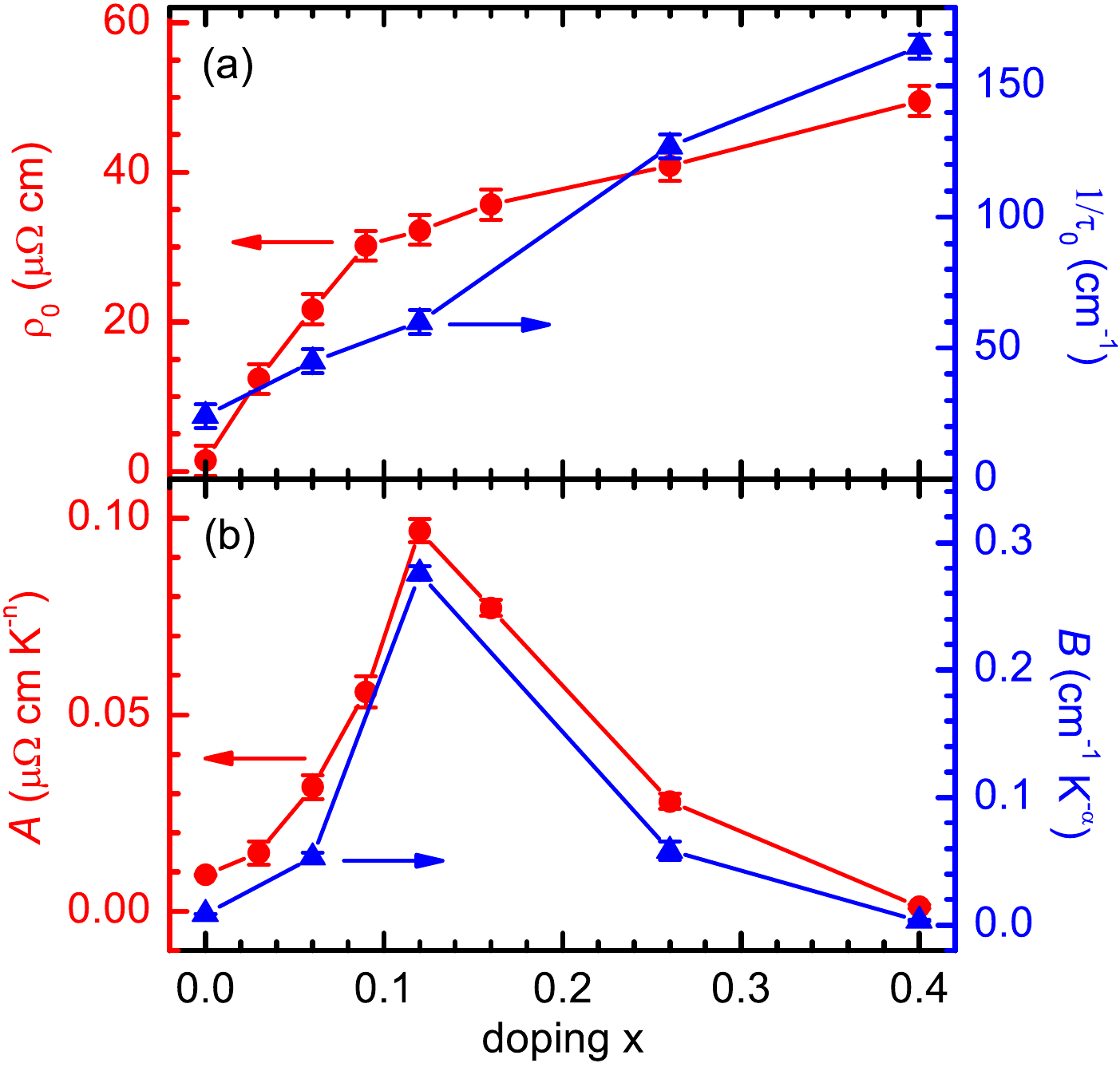}
\caption{ (color online) (a) Doping dependence of $\rho_{0}$ (solid circles) and $1/\tau_{0}$ (solid triangles), respectively. (b) Coefficients $A$ for transport (solid circles) and $B$ for optics (solid triangles) determined from the single power law fit as a function of Co concentration.}
\label{FigSRTvsO}
\end{figure}
%
\begin{figure}[tb]
\includegraphics[width=\columnwidth]{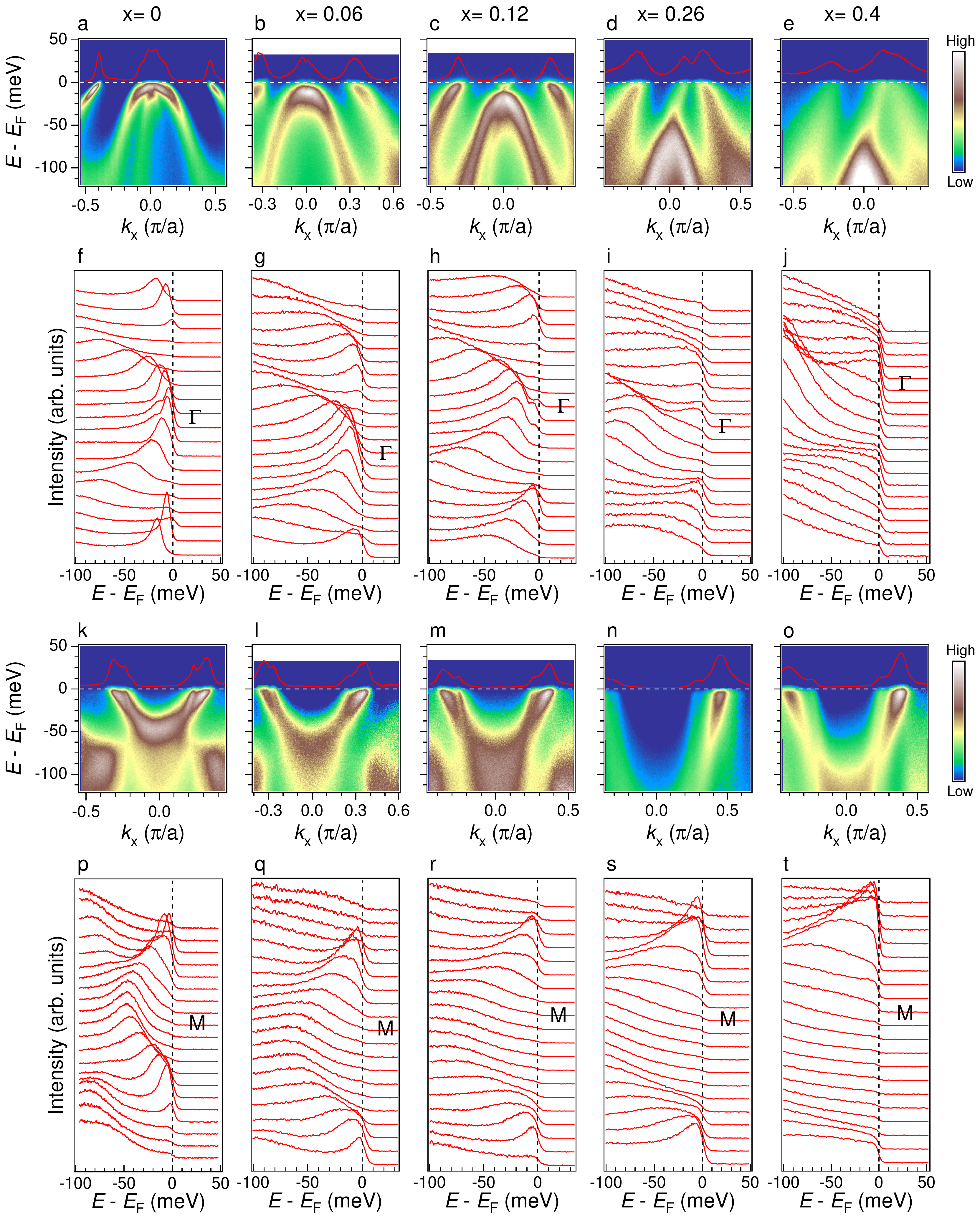}
\caption{ (color online) (a)--(j) ARPES intensity plots and corresponding EDCs at the Brillouin zone center. (k)--(t) ARPES intensity plots and corresponding EDCs at the $M$ point. All data are taken at 20~K along the $\Gamma$--$M$ direction. The red curves shown in the intensity plots are MDCs at the Fermi level.}
\label{FigSARPES}
\end{figure}
%
\begin{figure}[tb]
\includegraphics[width=0.95\columnwidth]{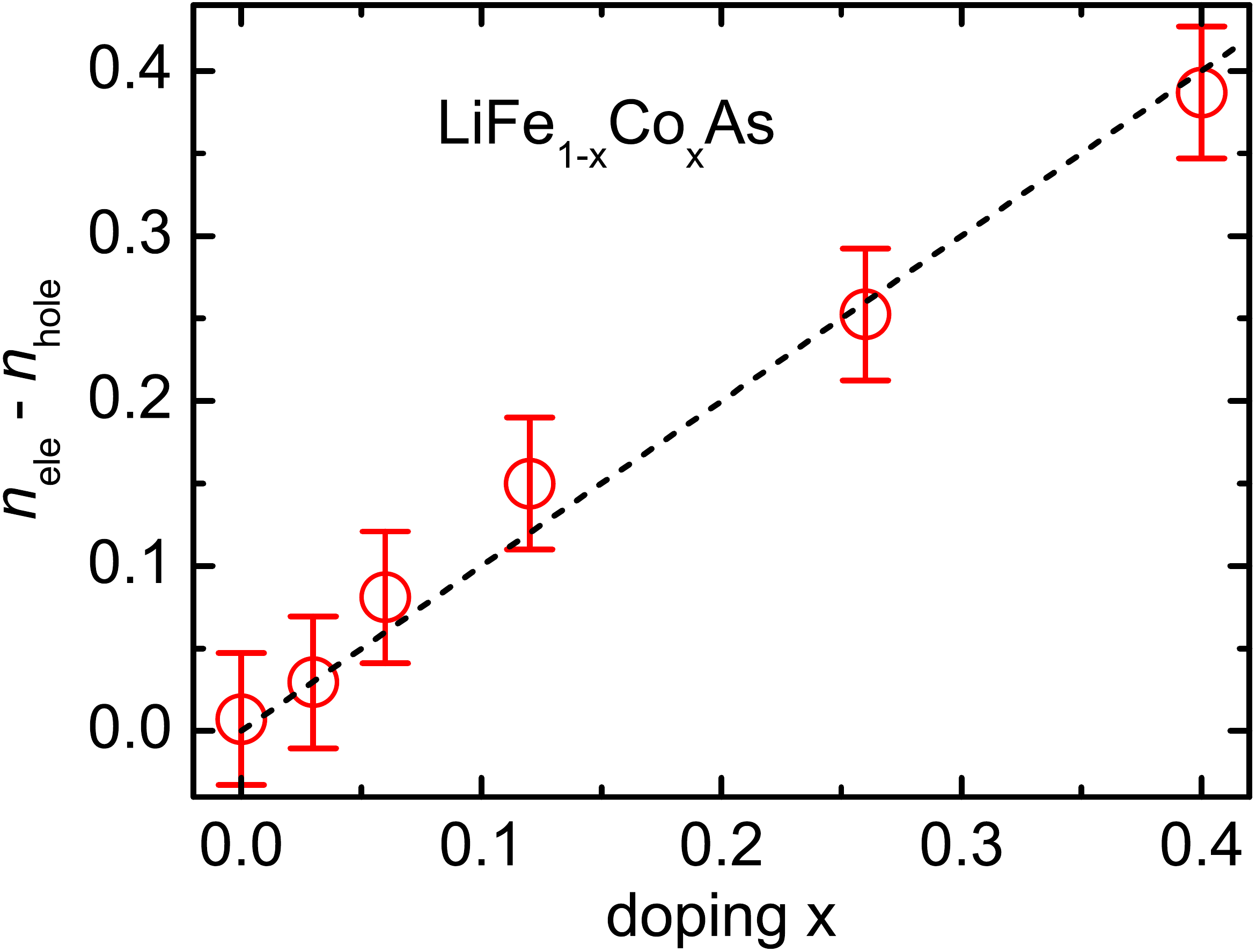}
\caption{ (color online) Total electronic carrier concentration (open circles) determined from the total algebraic Fermi surface volume at each Co concentration $x$. The dashed line corresponds to the carrier concentration assumed by adding one extra electron for each Fe substituted by Co.}
\label{FigSFSvol}
\end{figure}
\section{APPENDIX B: TRANSPORT}
\noindent\textbf{Resistivity measurements}\hspace{2mm}
The electrical transport measurements of \LFCA\ were carried out in a commercial physical properties
measurement system (PPMS) using the four-probe method. To prevent sample degradation, the electrical
contacts were prepared in a glove box and then the sample was protected by \emph{n}-grease before transferring
to the PPMS. Each sample was cut into a rectangular piece, so that its dimensions could be measured more accurately
with a microscope. With these precisely-measured geometry factors, the resistivity can be easily calculated
from the measured resistance. The resistivity determined from the transport measurements is then compared with
the values determined from the optical conductivity to ensure the consistency between different techniques. \\

\noindent\textbf{Single power law fitting}\hspace{2mm}
Figure~\ref{FigSRT} displays the resistivity as a function of temperature $\rho(T)$ (open circles) up
to 70~K for all 8 samples. For each substitution, $\rho(T)$ is fit to a single power law expression,
$\rho(T) = \rho_{0} + AT^{n}$, from $\sim$2~K (or $T_c$, whichever is greater) up to 70~K. The solid lines
through the data in each panel denote the fitting results. The power $n$, determined from the fitting, is
shown for the stoichiometric material and all the Co substitutions in the corresponding panels; the crossover
behavior of $n$ can be seen clearly.

In order to present the power-law behavior of the $\rho(T)$ curve more clearly, we plot $\rho(T)$ as a
function of $T^{n}$ for each substitution in Fig.~\ref{FigSRTn}, where $n$ is the power determined from the
single power law fitting. In this case, all the $\rho(T)$ curves can be perfectly described by linear
behavior, shown by the solid line in each panel.

\section{APPENDIX C: OPTICAL SPECTROSCOPY}
\noindent\textbf{Reflectivity}\hspace{2mm}
The temperature dependence of the absolute reflectivity $R(\omega)$ of \LFCA\ has been measured at a
near-normal angle of incidence for the stoichiometric material and 4 representative substitutions
using an \emph{in situ} overcoating technique~\cite{Homes1993a}.  For each sample, data were collected
at 17 different temperatures from 5~K to room temperature over a wide frequency range ($\sim$2~meV to
4~eV) on a freshly cleaved surface.  Because of the air-sensitive nature of the \LFCA\ samples, the
sample mounting and cleaving were done in a glove bag purged with high-purity Ar gas.  Immediately
after the cleaving, the sample was transferred to the vacuum shroud (also purged with Ar) with
the protection of a small Ar-purged plastic bag. The reproducibility of the experimental results was
checked by repeating the $R(\omega)$ measurements 2 or 3 times for each doping. Figure~\ref{FigSOpt}(a)--\ref{FigSOpt}(e) show $R(\omega)$
in the far-infrared region at 4 selected temperatures for 5 different Co concentrations.  For all the materials,
$R(\omega)$ approaches unity at zero frequency and increases upon cooling, indicating a metallic response. \\

%
%
\noindent\textbf{Kramers-Kronig analysis}\hspace{2mm}
The real part of the complex optical conductivity $\sigma_{1}(\omega)$ is determined from a
Kramers-Kronig analysis of the reflectivity.  Given the metallic nature of the
\LFCA\ materials, the Hagen-Rubens form $\left[ R(\omega) = 1 - A\sqrt{\omega}\, \right]$ is used
for the low-frequency extrapolation, where $A$ is chosen to match the data at the lowest-measured
frequency. Above the highest-measured frequency, $R(\omega)$ is assumed to be constant up to
1.0~$\times$~10$^{5}$~cm$^{-1}$, above which a free-electron response $\left[ R(\omega) \propto
\omega^{-4}\right]$ was used. \\

\noindent\textbf{Optical conductivity}\hspace{2mm}
Figure~\ref{FigSOpt}(f)--\ref{FigSOpt}(j) display $\sigma_{1}(\omega)$ at 4 selected temperatures for the stoichiometric
material and 4 different Co concentrations. The metallic behavior of these materials can be recognized
by the pronounced Drude-like peak centered at zero frequency. The zero-frequency value of
$\sigma_{1}(\omega)$ represents the dc conductivity $\sigma_{dc}$ which is in good agreement
with the values determined from transport measurements; the width of the Drude peak at half maximum
yields the quasiparticle scattering rate. As the temperature decreases, $\sigma_{dc}$ increases and
the Drude peak narrows. This indicates that the quasiparticle scattering rate decreases upon cooling,
dominating the temperature dependence of the electrical transport properties. \\

%
%
\noindent\textbf{Quasiparticle scattering rate}\hspace{2mm}
Figure~\ref{FigSOpt}(k)--\ref{FigSOpt}(o) show the quasiparticle scattering rate of the coherent narrow Drude component $1/\tau_{\rm co}$ as a
function of $T^{\alpha}$, where $\alpha$ is the power determined from the single power law
fit to $1/\tau_{\rm co}$ as a function of $T$.  Linear behavior can be clearly observed in each
panel as guided by the straight solid lines.\\

%
%
\noindent\textbf{Comparison between transport and optics}\hspace{2mm}
Figure~\ref{FigSRTvsO}(a) compares the doping dependence of $\rho_{0}$ (solid circles) and $1/\tau_{0}$ (solid triangles), where $\rho_{0}$ is determined by fitting $\rho(T)$ to $\rho(T) = \rho_{0} + A T^{n}$, and $1/\tau_{0}$ is derived by fitting $1/\tau_{\rm co}(T)$ to $1/\tau_{\rm co}(T) = 1/\tau_{0} + B T^{\alpha}$ for each doping. Both $\rho_{0}$ and $1/\tau_{0}$ grow as the Co concentration increases, indicating that impurities are introduced into the compounds by the Co substitution. The single power law fit also returns the coefficients $A$ for transport and $B$ for optics, respectively. As shown in Fig.~\ref{FigSRTvsO}(b), $A$ (solid circles) and $B$ (solid triangles) follow identical doping dependence, suggesting that the $T$ dependence in $\rho(T)$ and $1/\tau_{\rm co}(T)$ are governed by the same physics. Note that while the coefficients $A$ and $B$ follow exactly the same trace across doping, the detailed behaviors of $\rho_{0}$ and $1/\tau_{0}$ are slightly different. This is because $1/\tau_{0}$ is the residual scattering rate of the coherent narrow Drude component, which contributes to $\rho_{0}$ in parallel with the incoherent broad Drude component. However, as a common feature in all Fe-base superconductors~\cite{Wu2010a,Dai2013}, the $T$ dependence of the transport properties is dominated by the coherent narrow Drude component. Since $A$ and $B$ are the prefactors that describe the properties of the $T$ dependence in $\rho(T)$ and $1/\tau_{\rm co}(T)$, respectively, it is natural to expect similar behaviors for $A$ and $B$ if they are governed by the same physics.

\section{APPENDIX D: ANGLE-RESOLVED PHOTOEMISSION SPECTROSCOPY}
\noindent\textbf{Measurements}\hspace{2mm}
ARPES measurements were performed with a high-flux He discharge lamp. The energy
resolution was set at 12~meV and 3~meV for the Fermi surface mapping and high resolution measurements, respectively. The angular resolution was set at 0.2$^{\circ}$. Fresh surfaces for the ARPES measurements were obtained by an \emph{in situ} cleavage of the crystals
in a working vacuum better than 4$\times$10$^{-11}$~Torr. The Fermi energy (E$_{F}$) of the samples was
referenced to that of a gold film evaporated onto the sample holder.\\

\noindent\textbf{\boldmath ARPES intensity plots and energy/momentum distribution curves along the $\Gamma$--$M$ direction \unboldmath}\hspace{2mm}
ARPES intensity plots across the $\Gamma$ point for five representative dopings and their corresponding energy distribution curves (EDCs) are shown in Fig.~\ref{FigSARPES}(a)--\ref{FigSARPES}(e) and Fig.~\ref{FigSARPES}(f)--\ref{FigSARPES}(j), respectively. The same plots but crossing the $M$ point for each doping are shown in Fig.~\ref{FigSARPES}(k)--\ref{FigSARPES}(o) and Fig.~\ref{FigSARPES}(p)--\ref{FigSARPES}(t), respectively. The red curves in Fig.~\ref{FigSARPES}(a)--\ref{FigSARPES}(e) and Fig.~\ref{FigSARPES}(k)--\ref{FigSARPES}(o) are the momentum distribution curves (MDCs) at the Fermi level. The peak positions on MDCs correspond to the $k_{F}$ positions along the high symmetry line. Since the EDC and MDC peaks, as well as the band dispersions, are well defined at all dopings, the $k_{F}$ positions can be accurately determined for each Co concentration. The uncertainty of the $k_{F}$ position is mainly from the energy and momentum resolutions of our system settings. For these measurements the angular and energy resolutions of the system were set at 0.2$^{\circ}$ and 3~meV, respectively, which lead to a typical uncertainty of $\pm 0.01$~$\pi/a$ for $k_{F}$ at 21.2~eV photon energy.\\

\noindent\textbf{Evolution of the Fermi surface volume with doping}\hspace{2mm}
The evolution of the Fermi surface volume with doping is controlled by the Luttinger theorem: the total algebraic Fermi surface volume is directly proportional to the carrier concentration. We checked that this is the case in our study and we plot the results in Fig.~\ref{FigSFSvol}, which confirm that the total volume of the Fermi surface (open circles) satisfies the Luttinger theorem at each doping if we assume a rigid chemical potential shift caused by the introduction of one additional electron carrier per Fe atom substituted by Co (dashed line), as also observed theoretically~\cite{Berlijn2012} and experimentally~\cite{Neupane2011,Ideta2013} for the 122 family of iron pnictides.

\section{APPENDIX E: NUCLEAR MAGNETIC RESONANCE}
%
\begin{figure}[tb]
\includegraphics[width=\columnwidth]{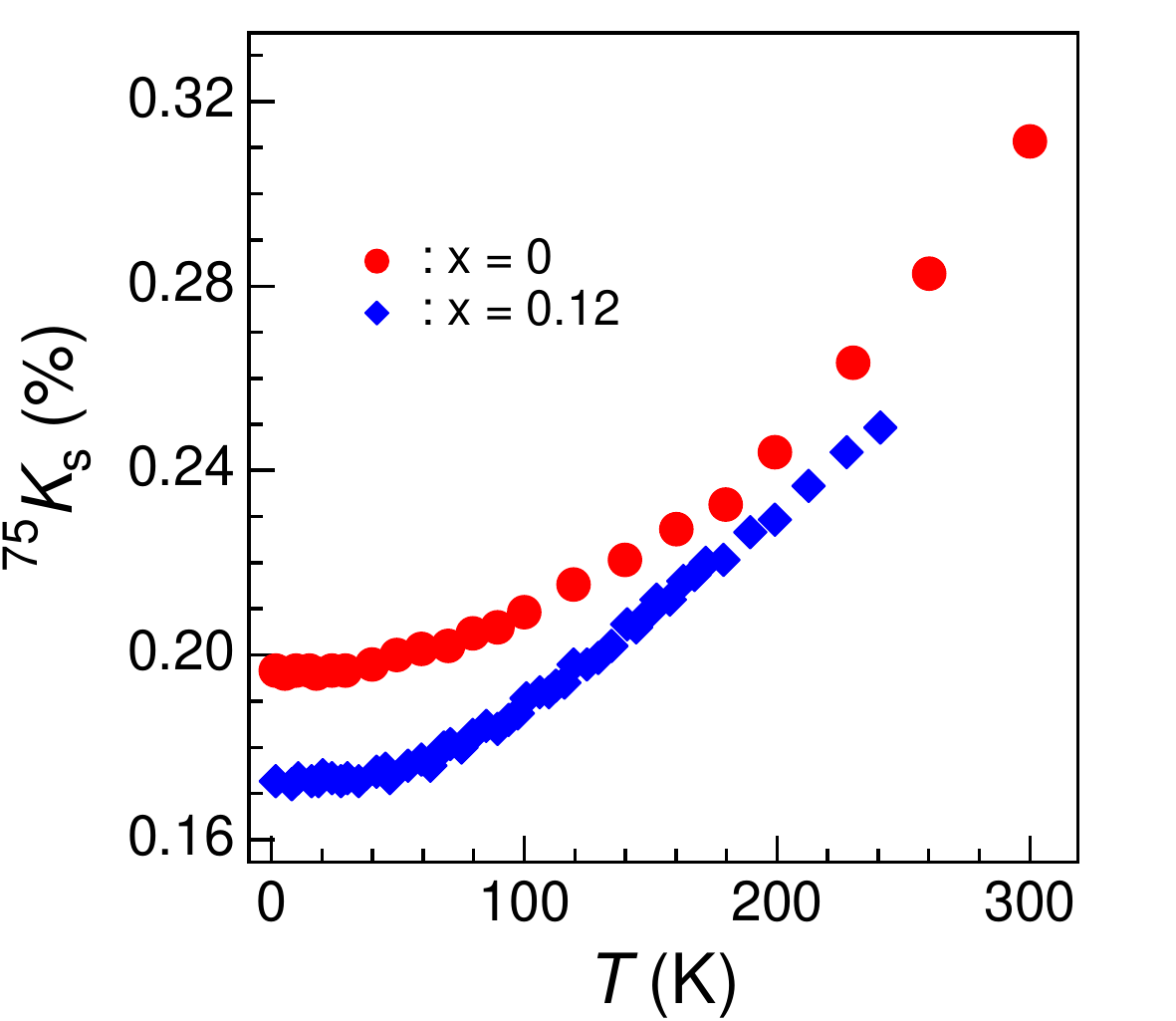}
\caption{ (color online) The Knight shifts of LiFe$_{1-x}$Co$_x$As for the $x=0$ and $x=0.12$ samples.
The Knight shift for both samples are measured with the external field perpendicular
to the \emph{a-b} plane to avoid calculation of the second
order quadruple correction. Though the size of the Knight shift along different directions may be
different, they share the same temperature dependence. Therefore we are able to take the $K_{s}$ at
$H\parallel{c}$ to estimate the behavior of $H\parallel{ab}$~\cite{Baek2013}.}
\label{FigSNMR}
\end{figure}
\noindent\textbf{Measurements}\hspace{2mm}
We perform the $^{75}$As nuclear magnetic resonance (NMR) measurements with the external field parallel
to the \emph{a-b} plane. The spin-lattice relaxation rate is measured by inversion-recovery method on
the central transition and the recovery curve is fit with a standard double exponential form for
an S=3/2 spin
%
%
\begin{equation}
  1-\frac{m(t)}{m(0)} = 0.9\exp{\left(\frac{-6t}{T_{1}}\right)} +
                        0.1\exp{\left(\frac{-t}{T_{1}}\right)} .
\label{DEForm}
\end{equation}
\\
%

%
%
\noindent\textbf{Knight shift}\hspace{2mm}
Figure~\ref{FigSNMR} shows the Knight shifts of $x=0$ and $x=0.12$ samples, respectively. The Knight
shift, which measures the uniform susceptibility, becomes temperature independent below 30~K,
indicating the low-temperature upturn in $1/T_1T$ comes from large-momentum spin fluctuations.
Note that since the NMR form factor for As is known to be broadly distributed in momentum
space in iron-based superconductors~\cite{Kitagawa2010}, both commensurate (close to FS nesting)
and incommensurate (away from nesting) low-energy spin fluctuations are captured by the
spin-lattice relaxation rate $1/T_1 T$.

%
%

\end{document}